\documentclass[prb,12pt,a4paper,showpacs]{revtex4}
\usepackage{amssymb}
\usepackage{amsfonts}
\usepackage{amsthm}

\theoremstyle{remark}
\newtheorem{defin}{Definition}
\newtheorem{lemma}{Lemma}
\newtheorem*{propo}{Proposition}
\theoremstyle{definition}
\newtheorem*{theor}{Theorem}
\newtheorem*{examp}{Example}

\def\fix{{\mathrm{Fix}}}
\def\tr{{\mathrm{tr}}}
\def\d{{\bar{\partial}}}
\def\dad{{\d^\dagger}}
\def\lap{{\Delta^{r,s}_\d}}
\def\R{{\mathbb{R}}}
\def\Rp{{\R^{+}}}
\def\Np{{\mathbb{N}^{+}}}
\def\C{{\mathbb{C}}}
\def\xC{{\otimes_\R \C}}
\def\P{{\mathbb{P}}}
\def\h{{\mathrm{hol}}}
\def\deg{{\mathrm{deg}}}

\begin{document}

\title{A Lefschetz fixed point theorem in gravitational lensing}

\author{Marcus C. Werner}
\email{mcw36@ast.cam.ac.uk}
\affiliation{Institute of Astronomy, University of Cambridge, Madingley Road, Cambridge CB3 0HA, United Kingdom}
\date{16 May 2007}

\begin{abstract}
Topological invariants play an important r\^{o}le in the theory of gravitational lensing by constraining the image number. Furthermore, it is known that, for certain lens models, the image magnifications $\mu_i$ obey invariants of the form $\sum_i \mu_i=1$. In this paper, we show that this magnification invariant is the holomorphic Lefschetz number of a suitably defined complexified lensing map, and hence a topological invariant. We also provide a heat kernel proof of the holomorphic Lefschetz fixed point formula which is central to this argument, based on Kotake's proof of the more general Atiyah-Bott theorem. Finally, we present a new astronomically motivated lens model for which this invariant holds.
\\
\textit{Version 2. Copyright 2007 American Institute of Physics. This article may be downloaded for personal use only. Any other use requires prior permission of the author and the American Institute of Physics. The following article appeared in J. Math. Phys. \textbf{48}, 052501 (2007) and may be found at} http://link.aip.org/?JMP/48/052501.

\end{abstract}

\maketitle

\section{Introduction}

Gravitational lensing provided one of the first successful tests of General Relativity\cite{Ed19} and proves an important tool in modern astronomy. Its theoretical treatment has two main inputs: the geometrical optics approximation and the static, scalar linearization of General Relativity. Within this framework, a deflector mass (e.g., a galaxy) acting as a lens can give rise to multiple images of a source (e.g., a background quasar) depending on the position of the source relative to caustics.
\\
It has emerged that the sum of signed magnifications $\mu(q_i)$ of the images $q_i$ may be a constant for some lens models, provided only that the source is in a caustic domain where the maximum number of images for the given model is attained, called the \textit{maximal domain}\cite{Rh97}. This remarkable property is known as a \textit{magnification invariant}. The first instance of such an invariant was found by Witt and Mao\cite{Wi95} for a binary point mass as lens which can produce up to five images,
\[
\sum_{i=1}^5 \mu (q_i) =1.
\]
This result was generalized\cite{Rh97} and shown to hold also for the Plummer model\cite{We06}. Constant higher moments\cite{Wi00}, called \textit{configuration invariants}\cite{Hu01}, and further examples of magnification invariants have been found. Powerful methods to derive such invariants using one-dimensional and multidimensional complex analysis were developed by Evans and Hunter\cite{Ev02,Hu01} and Dalal and Rabin\cite{Da00}, respectively. Dalal and Rabin\cite{Da00} also found an explanation for these invariants in terms of the Global Residue Theorem, noting that their properties do not suggest a topological invariant since general models do not obey magnification invariants.
\\
Revisiting this problem, we shall consider images as fixed points of a suitably defined complex rational lensing map. It turns out that, from this point of view, the complex magnification sum can be understood as a topological invariant, namely the holomorphic Lefschetz number induced by this map.
\\
The holomorphic Lefschetz fixed point formula in our argument is a special case of the Atiyah-Bott theorem\cite{At66}, and a proof using characteristic classes was given by Patodi\cite{Pa73}. In section II, we sketch a heat kernel proof of the holomorphic Lefschetz formula based on Kotake's proof\cite{Ko69} of the full Atiyah-Bott theorem. In particular, we specialize it to K\"{a}hler manifolds and hence apply Hodge decomposition instead of a more general lemma used by Kotake. The application to gravitational lensing is then presented as Theorem in section III. Regarded as a consequence of the Atiyah-Bott theorem, this furnishes a topological explanation for magnification invariants in gravitational lensing provided that all images in the maximal domain are real. This condition is further discussed in section IV where we also provide a new example of this magnification invariant in the astronomical context by considering a simple model for a filament of galaxies. 
\\
Finally, we note for clarity that Petters and Wicklin\cite{Pe98} have introduced a notion of fixed points in gravitational lensing which is different from the one considered here. They defined fixed points as image positions coinciding with the source position, whereas here all images are fixed points of a suitably constructed map.

\section{Fixed point theory}
\subsection{Preliminaries}

We briefly review aspects of complex manifolds and introduce notation as relevant to the present work. A fuller discussion can be found in Griffiths and Harris\cite{Gr78}.
\\
Let $(M,g)$ be a closed K\"{a}hler manifold with $\dim_\C M=m$ and charts with local holomorphic coordinates $z_{\mu}=x_{\mu}+iy_{\mu}, \ 1 \leq \mu \leq m$. Recall that the complexified cotangent space at $p \in M$ of the underlying differentiable manifold $M$ of real dimension $2m$ at can be decomposed into its holomorphic and antiholomorphic components $T^\ast_pM \xC = T^{\ast 1,0}_p M \oplus T^{\ast 0,1}_p M$ spanned by $\langle dz_\mu \rangle=T^{\ast 1,0}_pM, \ \langle d\bar{z}_\mu\rangle=T^{\ast 0,1}_p M$, respectively. Similarly, differential $k$-forms may be complexified and uniquely decomposed into bidegree $(r,s)$ according to
\begin{equation}
\Omega^k(M) \xC =\bigoplus_{r+s=k}\Omega^{r,s}(M) 
\label{complex1}
\end{equation}
where $1 \leq r,s \leq m$. Likewise, the exterior derivative $d$ is a real operator and can be separated into the Dolbeault operators thus
\[
d=\partial+\d \ \ \mbox{where} \ \ \partial: \Omega^{r,s}(M) \rightarrow \Omega^{r+1,s}(M), \ \d: \Omega^{r,s}(M) \rightarrow \Omega^{r,s+1}(M)
\]
which satisfy the usual nil-potency conditions. Furthermore, since $M$ is K\"{a}hler, it is in particular Hermitian whence the metric components $g_{\mu\nu}=0=g_{\bar{\mu}\bar{\nu}}$. Then the complex conjugate Hodge dual can be defined,
\[
\bar{\ast} : \Omega^{r,s}(M) \rightarrow \Omega^{m-r,m-s}(M) \ \ \mbox{such that} \ \ \bar{\ast}\omega=\overline{\ast \omega}
\]
the inner product of two $(r,s)$-forms $\phi, \psi$,
\[
(\phi,\psi) = \int_M \phi \wedge \bar{\ast} \psi,
\]
and also a nil-potent differential operator $\dad$ adjoint to the Dolbeault operator $\d$ with respect to this inner product,
\begin{equation}
\dad: \Omega^{r,s}(M) \rightarrow \Omega^{r,s-1}(M), \ \ \mbox{such that} \ \ (\phi, \d\psi)=(\dad\phi,\psi).
\label{diff2}
\end{equation}
In addition to the real Laplacian $\Delta^k_d:\Omega^k(M)\rightarrow \Omega^k(M)$, we can now define the $\d$-Laplacian induced by the Hermitian metric,
\begin{equation}
\lap=\d \dad +\dad \d : \Omega^{r,s}(M) \rightarrow \Omega^{r,s}(M)
\label{laplace}
\end{equation}
which commutes with $\d$ and $\dad$ on account of their nil-potency property. Elements of its kernel $\mathrm{ker} \ \lap(M)$ are called $\d$-harmonic forms. Note that the K\"{a}hler property may hence be expressed as 
\begin{equation}
\Delta_d^k=2\lap
\label{kahler}
\end{equation}
where $r+s=k$. Any form $\omega$ of bidgree $(r,s)$ may according to Hodge be uniquely and orthogonally decomposed into a sum of $\d$-exact, $\dad$-exact and $\d$-harmonic forms which in terms of Green's operator $(\lap)^{-1}$ and the harmonic projection operator $P^{r,s}: \Omega^{r,s}(M)\rightarrow \mathrm{ker}\ \lap(M)$ may be realized by
\begin{equation}
\omega=\d(\dad(\lap)^{-1}\omega)+\dad(\d(\lap)^{-1}\omega)+P^{r,s}\omega.
\label{hodge2}
\end{equation}
The $(r,s)$-cohomology spaces are defined as usual by
\[
H^{r,s}_\d =\mathrm{ker} \ \d (\Omega^{r,s}(M)) / \mathrm{im} \ \d (\Omega^{r,s-1}(M)).
\]
Elements of their orthonormal bases are denoted by $[\epsilon^{r,s}_i], \ 1\leq i\leq h^{r,s}$ where the Hodge numbers $h^{r,s}(M)=\dim_\C H^{r,s}_\d(M)$ are their complex dimensions.
\\
Hodge decomposition (\ref{hodge2}) implies that there is a unique $\d$-harmonic representative for each $[\epsilon^{r,s}_i]$. On the other hand, since $\d$-exact forms are orthogonal to $\d$-harmonic forms, they represent non-trivial equivalence classes. Hence we have Hodge's theorem
\[
H^{r,s}_\d(M) \cong \mathrm{ker}\ \lap(M).
\]
Consider how the $(r,s)$-cohomology spaces $H^{r,s}_\d(M)$ are related to the complexified de Rham cohomology spaces $H^k(M) \xC$. The K\"{a}hler property (\ref{kahler}) implies that $\Delta^k_d \omega=0 \Leftrightarrow \lap \omega=0$, and therefore, by Hodge's theorem, the complexified cohomology spaces simply decompose into $(r,s)$-cohomology according (\ref{complex1}),
\begin{equation}
H^k(M) \xC=\bigoplus_{r+s=k}H^{r,s}_\d(M), \ \ \mbox{whence} \ \ b^k(M)=\sum_{r+s=k}h^{r,s}(M)
\label{dolhomo}
\end{equation}
because the Betti numbers $b^k(M)=\dim_\R H^k(M)=\dim_\C H^k(M) \xC$. Moreover, an $(r,s)$-form is $\d$-harmonic iff its complex conjugate $(s,r)$-form is $\d$-harmonic on the K\"{a}hler manifold $M$ since $\Delta^k_d$ is a real operator. Furthermore, Poincar\'{e} duality implies that $h^{r,s}(M)=h^{m-r,m-s}(M)$.
\\
Now the smooth map $f: M \rightarrow M $ induces the usual linear pull-back map $f^\ast$ and push-forward map $f_\ast$ on the cotangent and tangent spaces, respectively. The Jacobian matrix $J_f$ of $f$ is then defined by the action on basis vectors of $T_pM$ where $f_\ast: \mathbf{e}_p \mapsto \mathbf{e}_{f(p)}$ such that $f_\ast \mathbf{e}_p =J_f \mathbf{e}_{f(p)}$.
\\
A fixed point $q \in M$ of $f$ satisfies $f(q)=q$, and the set of fixed points is denoted by $\fix(f)=\Delta \cap \Gamma_f$ where $\Delta=\{ (p,p) \ | \ p \in M \} \subset M \times M$ is the diagonal, and $\Gamma_f=\{ (p,f(p)) \ | \ p \in M\} \subset M \times M$ is the graph of $f$. Note that compactness of $M$ ensures that $\fix(f)$ is finite. We will be concerned with fixed points of $f$ that are isolated and non-degenerate so that $f$ satisfies the transversality condition $\Gamma_f \pitchfork \Delta$, that is,
\begin{equation}
T_{(q,q)}\Gamma_f+T_{(q,q)}\Delta=T_{(q,q)}(M\times M) \ \forall \ q \in \fix(f).
\label{trans}
\end{equation}
Finally, let $I_m$ denote the $m$-dimensional identity.

\subsection{Holomorphic Lefschetz formula}
The holomorphic Lefschetz number is a topological invariant for a geometric endomorphism which can be defined if $f$ is holomorphic since, in this case, the pull-back induced by $f$ preserves the bidegree of complex forms.

\begin{defin}
Let $f: M \rightarrow M$ be holomorphic. Then the induced map $\gamma^{r,s}: \Omega^{r,s}(M) \rightarrow \Omega^{r,s}(M)$ defined by $\gamma^{r,s} \omega(p)=f^\ast \omega(f(p)),\  p \in M$ such that $\gamma^{r,s+1}\d=\d \gamma^{r,s}$ is called geometric endomorphism.
\label{definendo}
\end{defin}

Note also that complexification of the Jacobian of a holomorphic map $f$ respects the decomposition into holomorphic and antiholomorphic components,
\begin{equation}
J_f \xC=J^{1,0}_f \oplus J^{0,1}_f
\label{jac2}
\end{equation}
where $J^{1,0}_f$ and $J^{0,1}_f=\overline{J^{1,0}_f}$ act on $T^{1,0}_{f(p)}M$ and $T^{0,1}_{f(p)}M$, respectively.
\\
The commutation property in Definition \ref{definendo} implies that the geometric endomorphism on $(r,s)$-forms is also an endomorphism on $(r,s)$-cohomology spaces. This topological property allows a global definition for the holomorphic Lefschetz number of $M$ induced by $f$.

\begin{defin}
Let $f: M \rightarrow M$ be holomorphic, inducing the geometric endomorphism $\gamma^{r,s}$. Then the holomorphic Lefschetz number is given globally by
\[
L_\h(f)=\sum_{s=0}^{m} (-1)^s \ \tr \ \gamma^{0,s} (H^{0,s}_\d(M)).
\]
\label{definlefs}
\end{defin}

The holomorphic Lefschetz formula connects this global definition with local properties of $\fix(f)$. Now Hodge's theorem shows that the kernel of the Laplacian provides information on the topology of the manifold. Hence we shall use heat kernel methods to establish this connection.

\begin{defin}
Let $\omega(t,p) \in \Rp \times \Omega^{0,s}(M)$ be a solution of the Dolbeault heat equation $(\frac{\partial}{\partial t}+\lap)\omega(t,p)=0$ where $p \in M$, the time $t\in \Rp$ is the flow parameter and the initial condition is $\omega(p) \in \Omega^{0,s}(M)$. Then the heat operator $K^{0,s}(t): \Omega^{0,s}(M) \rightarrow \Rp\times\Omega^{0,s}(M)$ and the heat kernel double form $\kappa^{0,s}(t,p,p') \in \Rp\times\Omega^{0,s}(M)\otimes \Omega^{0,s}(M)$ are defined by
\[
\omega(t,p)= K^{0,s}(t)\omega(p)=\int_{M} \kappa^{0,s}(t,p,p') \wedge \bar{\ast} \ \omega(p').
\]
\label{definheat}
\end{defin}

 Note also that the integral of the pointwise trace of a double form $\upsilon(p,p')=\phi(p)\otimes\psi(p')$ where $\omega(p),\psi(p')\in \Omega^{0,s}(M)$ is given by the inner product
\begin{equation}
\int_M\tr \ \upsilon(p,p)=(\phi,\psi).
\label{inner2}
\end{equation}

\begin{lemma}
Let $f: M \rightarrow M$ be holomorphic and $\kappa^{0,s}$ the heat kernel of Definition \ref{definheat}. Then the holomorphic Lefschetz number may be written
\[
L_\h(f)=\sum_{s=0}^{m}(-1)^s\int_M\tr \ f^{\ast 0,s}\kappa^{0,s}(t,f(p),p)
\]
\label{lemmalefs1}
\end{lemma}
\begin{proof}
Firstly, introduce the operator
\[
S^{0,s}(t)=K^{0,s-1}(t)\dad (\Delta_\d^{0,s})^{-1}:\Omega^{0,s}(M) \rightarrow \Rp \times \Omega^{0,s-1}(M),
\]
and note that both the heat operator and Green's operator commute with $\d$ since $\lap$ does. Furthermore, since $P^{0,s} \omega$ and $K^{0,s}(t)P^{0,s}\omega$ are both solutions of the heat equation for $t \in \Rp$ and have the same initial value for any $\omega$, they are identical by the uniqueness of solutions of the heat equation. We can now apply the Hodge decomposition (\ref{hodge2}) to the initial solution in Definition \ref{definheat} to obtain an operator equation,
\begin{equation}
K^{0,s}(t)=\d S^{0,s}(t)+S^{0,s+1}(t)\d+ P^{0,s}.
\label{heat3}
\end{equation}
Because of the orthogonality of the Hodge decomposition, this may be equivalently be expressed in terms of the kernels of the operators of (\ref{heat3}). Furthermore, Hodge's theorem implies that the harmonic projection operator is a map $P^{0,s}: \Omega^{0,s}(M) \rightarrow H_\d^{0,s}(M)$. Therefore its kernel $\pi^{0,s}$, defined analogously to $\kappa^{0,s}$ in Definition \ref{definheat}, may be written
\begin{equation}
\pi^{0,s}(p,p')=\sum_{i,j}\epsilon^{0,s}_i(p)\otimes\epsilon^{0,s}_j(p')
\label{pkernel}
\end{equation}
in terms of the orthonormal basis for the $(0,s)$-cohomology spaces. Using this basis to express the holomorphic Lefschetz number, one can apply (\ref{pkernel}) and (\ref{heat3}) in terms of kernels. But since the kernels of $S^{0,s}(t)$ cancel due to the alternating sum in $L_\h(f)$, one finds
\begin{eqnarray*}
L_\h(f)&=&\sum_{s=0}^{m}(-1)^s\sum_{i=1}^{h^{0,s}}(\gamma^{0,s}\epsilon^{0,s}_i, \epsilon^{0,s}_i)=\sum_{s=0}^{m}(-1)^s\int_M\tr \ \gamma^{0,s}\pi^{0,s}(p,p')|_{p=p'}\nonumber \\
&=&\sum_{s=0}^{m}(-1)^s\int_M\tr \ \gamma^{0,s}\kappa^{0,s}(t,p,p')|_{p=p'}=\sum_{s=0}^{m}(-1)^s\int_M\tr \ f^{\ast 0,s}\kappa^{0,s}(t,f(p),p)
\end{eqnarray*}
using Definition \ref{definendo} of the geometric endomorphism. This is the required result.
\end{proof}

\begin{propo}[Holomorphic Lefschetz formula]
Let $f:M\rightarrow M$ be holomorphic and transversal in open neighborhoods $U_q\ni q  \ \forall q \in \fix{f}$. Then the holomorphic Lefschetz number is given locally by,
\[
L_\h(f)=\sum_{q \in \fix(f)}\frac{1}{\det(I_m-J^{1,0}_f)(q)}.
\]
\label{propolefs}
\end{propo}
\begin{proof}
The K\"{a}hler property (\ref{kahler}) implies that solutions of the Dolbeault heat equation of Definition \ref{definheat} are also solutions of the real heat equation with a rescaling of time. Assuming isolated fixed points $q \in \fix(f)$, we can approximate the metric $g$ on $M$ by a Euclidean metric on the $U_q$ for $t\rightarrow 0$. Then the real Laplacian on functions is given by $\Delta^0_d=-\nabla \cdot \nabla$ on $U_q$ with local coordinates $\mathbf{x}=(x^1,\ldots, x^m, y^1, \ldots, y^m) \in \R^{2m}$, where $\nabla=(\frac{\partial}{\partial x_1},\ldots, \frac{\partial}{\partial x_m},\frac{\partial}{\partial y_1},\ldots, \frac{\partial}{\partial y_m})$ and the dot product is with respect to the Euclidean metric. Hence the heat equation can be solved with a Fourier transform, yielding a local approximation for the heat kernel such that
\[
\kappa^{0,s}(t,\mathbf{x},\mathbf{x'})=\frac{1}{(2\pi)^{2m}}\int \exp\left(i(\mathbf{x}-\mathbf{x'})\cdot \mathbf{\xi}-t\mathbf{\xi}^2\right)d\xi^{2m} \phi(\mathbf{x}) \otimes \phi(\mathbf{x'}).
\]
where $\phi= d\bar{z}_{\bar{\mu}_1}\wedge \ldots \wedge d\bar{z}_{\bar{\mu}_s}$ denotes the basis elements on $\Omega^{0,s}(M)$ and $\mathbf{\xi} \in \R^{2m}$. Since $L_\h(f)$ does not depend on time by the global Definition \ref{definlefs}, we may take the limit $t\rightarrow 0$ to evaluate it. From the decomposition (\ref{jac2}), the pull-back $f^{\ast0,s}$ on $\Omega^{0,s}(U_q)$ is the $s$th exterior power of $J^{0,1}_f$. Then the trace operation (\ref{inner2}) provides the volume form on $M$, and hence the holomorphic Lefschetz number becomes
\[
L_\h(f)=\sum_{q \in \fix(f)}\sum^{m}_{s=0}\frac{(-1)^s}{(2\pi)^{2m}}\lim_{t\rightarrow 0}\int_{U_q}\int \tr \ \Lambda^s(J^{0,1}_f) \exp \left(-i(\mathbf{x}-f(\mathbf{x}))\cdot \mathbf{\xi}+t\mathbf{\xi}^2\right)d\xi^{2m} dx^{2m}.
\]
Define the local coordinate $\mathbf{u}=\mathbf{x}-f(\mathbf{x})$ on $U_q$. Then the integration measure transforms in the usual way since the fixed points are non-degenerate. Hence
\[
L_\h(f)=\sum_{q\in \fix(f)} \sum_{s=0}^{m} \frac{(-1)^s}{(2\pi)^{2m}} \int_{U_q}\int \frac{\tr \ \Lambda^s(J^{0,1}_f)(\mathbf{u}) e^{-i \mathbf{u}\cdot\mathbf{\xi}}}{|\det(I_{2m}-J_f)|(\mathbf{u})} d\xi^{2m} du^{2m}
\]
Recall $\int \exp (-i \mathbf{u} \cdot \mathbf{\xi})d\xi^{2m}/(2\pi)^{2m}=\delta(\mathbf{u})$ and $\mathbf{u}=\mathbf{0}$ selects the fixed point in $U_q$. Finally, using the equality $\sum_{s=0}^{m} (-1)^s \ \tr \ \Lambda^s(J^{0,1}_f)=\det(I_m-J^{0,1}_f)$ and noting that $\det(I_{2m}-J_f)=\det(I_{2m}-J_f\xC)=\det(I_m-J^{1,0}_f)\det(I_m-J^{0,1}_f)$, the result follows.
\end{proof}

\section{Gravitational lensing}
\subsection{Preliminaries}
The standard physical treatment of gravitational lensing, as indicated in the introduction, is the thin lens approximation. We will now briefly outline this concept, following Petters, Levine and Wambsganss\cite{Pe01}.
\\
Consider a point-like light source at $\mathbf{y}=(y_1,y_2) \in S = \R^2$, the source plane. A translucent gravitational lens is projected into the lens plane $L=\R^2$, with coordinates $\mathbf{x}=(x_1,x_2)$, which is parallel to $S$, and between $S$ and the observer. The projected surface density of the lens $\sigma: L \rightarrow \R$ gives rise to the deflection potential $\Psi$ via a Poisson equation,
\[
\Psi: L \rightarrow \R, \ \Delta^0_d \Psi=-2\sigma,
\]
which defines the lens model for a given model of the deflector mass. Physically, the deflection potential may be thought of as the Shapiro time delay for a trial path from $S$ to the observer. Another contribution to the time delay stems from the path length. The overall time delay may now be written $\Phi_\mathbf{y}(\mathbf{x})=(\mathbf{x}-\mathbf{y})^2/2-\Psi(\mathbf{x})$ and is called the Fermat potential. The actual ray path is determined by Fermat's principle which in this notation becomes $\nabla \Phi_{\mathbf{y}}(\mathbf{x})=0$. Therefore the lensing map $\eta$ becomes
\begin{equation} 
\eta: L \rightarrow S, \  \mathbf{x} \mapsto \mathbf{y}=\mathbf{x}-\nabla \Psi.
\label{lensmap}
\end{equation}
Thus for a given source position, real roots of the lensing equation (\ref{lensmap}) correspond to the images of a point source seen at $\mathbf{x} \in L$. Note also that $\nabla \Psi$ is the deflection angle projected into the lens plane.
\\
Source and lens plane can be complexified as follows,
\[
\zeta=y_1+iy_2 \in S\xC, \ z=x_1+ix_2 \in L\xC.
\]
Similarly, we introduce a complexified deflection angle,
\begin{equation}
\alpha=\frac{\partial \Psi}{\partial x_1} + i\frac{\partial \Psi}{\partial x_2}.
\label{potential}
\end{equation}
Hence the lensing map may be written
\begin{equation}
\zeta=z-\alpha(z,\bar{z}).
\label{lens2}
\end{equation}
Since in geometrical optics light rays are conserved, the signed image magnification $\mu$ is proportional to the solid angle of the ray bundle and is hence given by
\begin{equation}
\mu=\frac{1}{\det J_\eta}=\frac{1}{\det J_\eta\xC}.
\label{mag}
\end{equation}
A negative magnification is to be interpreted as an orientation inverting lens map.

\subsection{Complex magnification invariant}
In order to apply the holomorphic Lefschetz formula to gravitational lensing, we need to consider the geometric endomorphism induced by a holomorphic map. But the lens map $\eta$ is, in general, not holomorphic in the standard complexification since (\ref{lens2}) depends on both $z$ and $\bar{z}$. This problem can be circumvented by defining $z_1\equiv z, \ z_2 \equiv\bar{z}$ as independent holomorphic coordinates on $\C^2$.

\begin{defin}
A rational lensing map is a transversal map $f=(f_1,f_2): \C^2\rightarrow \C^2$ such that
\begin{eqnarray*}
f_1(z_1,z_2)&=&\zeta + \alpha_1(z_1,z_2) \nonumber \\
f_2(z_1,z_2)&=&\bar{\zeta}+\alpha_2(z_1,z_2), 
\end{eqnarray*}
where $\zeta, \bar{\zeta}$ are constant and the complexified deflection angle $\alpha_1=\alpha, \alpha_2=\bar{\alpha}$ from (\ref{potential}) is a rational map
\[
\alpha_\mu(z_1,z_2)=\frac{U_\mu(z_1,z_2)}{V_\mu(z_1,z_2)}=\frac{\sum_{i,j}u_{\mu ij}z_1^i z_2^j}{\sum_{k,l}v_{\mu kl}z_1^k z_2^l}
\]
for $1 \leq \mu \leq 2$, such that the polynomials $U_\mu,V_\mu$ are of degree $\deg U_\mu < \deg V_\mu$. $\fix(f)$ is called the set of complex images.
\label{definlens1}
\end{defin}

With this definition, $f$ is holomorphic since $\frac{\partial f_\mu}{\partial \bar{z}_\nu}\equiv 0, \ 1 \leq \mu,\nu \leq 2$. Note also that, by construction, $\fix(f)$ contains the real images of the lensing map $\eta$. Furthermore, the condition $\deg U_\mu < \deg V_\mu$ in Definition \ref{definlens1} is natural for finite lenses since it implies that the complexified deflection angle $\alpha_\mu \rightarrow 0$ as $z_\mu \rightarrow \infty$.
\\
However, $\C^2$ is not a closed K\"{a}hler manifold and so we need to compactify the complexified lens and source planes. Recall that complex projective space $\C\P^2=\left(\C^3 \backslash \mathbf{0}\right)/^\sim$, with local homogeneous coordinates $(Z_0,Z_1,Z_2)\sim(\lambda Z_0,\lambda Z_1,\lambda Z_2), \ \lambda \neq 0$, is K\"{a}hler with the Fubini-Study metric, and closed. We can write as usual,
\[
z_1=\frac{Z_1}{Z_0}, \ z_2=\frac{Z_2}{Z_0} \ \ \mbox{for} \ \ Z_0 \neq 0.
\]
Then $f$ induces a map on $\C\P^2$.

\begin{defin}
A projective lensing map $F$ induced by a rational lensing map $f$ is a map $F=(F_0:F_1:F_2): \C\P^2 \rightarrow \C\P^2$
\begin{eqnarray*}
F_0(Z_0,Z_1,Z_2)&=&Z_0 \nonumber \\
F_1(Z_0,Z_1,Z_2)&=&Z_0\zeta + A_1(Z_0,Z_1,Z_2) \\
F_2(Z_0,Z_1,Z_2)&=&Z_0\bar{\zeta}+A_2(Z_0,Z_1,Z_2)
\end{eqnarray*}
where, using Definition \ref{definlens1},
\[
A_\mu=Z_0^{\deg V_\mu-\deg U_\mu+1}\frac{\sum_{i,j}u_{\mu ij}Z_0^{\deg U_\mu-i-j}Z_1^i Z_2^j}{\sum_{k,l}v_{\mu kl}Z_0^{\deg V_\mu-k-l}Z_1^k Z_2^l}
\]
is homogeneous of degree one.
\label{definlens2}
\end{defin}

Hence on the restriction $\C^2: Z_0=1$ we recover $F|_{\C^2}=f$.

\begin{lemma}
Let $F$ be a projective lensing map. Then $L_\h(F)=1$.
\label{lemmalefs2}
\end{lemma}
\begin{proof}
By Definition \ref{definlens2}, $F: \C\P^2 \rightarrow \C\P^2$. We shall show that $L_\h(F)=h^{0,0}(\C\P^2)=1$. Recall from (\ref{dolhomo}) that for the K\"{a}hler manifold $\C\P^2$,
\[
H^k(\C\P^2) \xC=\bigoplus_{r+s=k}H^{r,s}_\d(\C\P^2), \ \ \mbox{so that} \ \ b^k(\C\P^2)=\sum_{r+s=k}h^{r,s}(\C\P^2).
\]
Given, furthermore, the symmetry and Poincar\'{e} duality property, the only independent Hodge numbers of $\C\P^2$ are $h^{0,0}, \ h^{0,1}, \ h^{1,1}, \ h^{0,2}$. Then using the Betti numbers of $\C\P^2$ we have
\begin{eqnarray*}
1&=&b^0(\C\P^2)=h^{0,0}(\C\P^2) \\
0&=&b^1(\C\P^2)=2\sum_{\stackrel{r+s=1}{r<s}}h^{r,s}(\C\P^2)=2h^{0,1}(\C\P^2) \\
1&=&b^2(\C\P^2)=\sum_{r+s=2}h^{r,s}(\C\P^2)=h^{1,1}(\C\P^2)+2h^{0,2}(\C\P^2) 
\end{eqnarray*}
whence $h^{0,0}=1, \ h^{0,1}=0, \ h^{1,1}=1, \ h^{0,2}=0$. Now in the global Definition \ref{definlefs} of the holomorphic Lefschetz number only the $H^{0,s}_\d$ contribute, and the result follows.
\end{proof}

\begin{lemma}
Let $F$ be a projective lensing map and $f$ be a rational lensing map. Then $\fix(F)=\fix(f)$.
\label{lemmalefs3}
\end{lemma}
\begin{proof}
The complex projective space may be decomposed in the usual way, $\C\P^2=\C^2 \cup \C\P^1$, where $\C^2: Z_0=1$ and $\C\P^1: Z_0=0$. Now Definition \ref{definlens2} implies that $A_\mu(0,Z_1,Z_2)=0$. Hence the restriction on $\C\P^1$ is given by $F|_{\C\P^1}=(F_1(Z_1,Z_2):F_2(Z_1,Z_2))=(0,0) \notin \C\P^1$. Therefore $\fix(F|_{\C\P^1})=\emptyset$, and the only fixed points of $F$ are those of $F|_{\C^2}=f$, as required.
\end{proof}

\begin{theor}
Let $f$ be a rational lensing map inducing the projective lensing map $F$. Then the sum of the magnifications $\mu (q)$ of the complex images $q$ obeys
\[
\sum_{q \in \fix(f)} \mu(q)=L_\h(F)=1.
\]
\label{theorlens}
\end{theor}
\begin{proof}
From Lemma \ref{lemmalefs3}, $\fix(F)=\fix(f)$. Hence using the holomorphic Lefschetz formula from the Proposition,
\[
L_\h(F)=\sum_{q \in \fix(f)}\frac{1}{\det(I_2-J^{1,0}_f)(q)}.
\]
Now an explicit calculation of the Jacobian of $f$ from Definition \ref{definlens1} shows that
\[
I_2-J^{1,0}_f=
\left(\begin{array}{cc}
1-\frac{\partial \alpha_1}{\partial z_1}|_{z_2} & \frac{\partial \alpha_1}{\partial z_2}|_{z_1} \\ 
\frac{\partial \alpha_2}{\partial z_1}|_{z_2} & 1-\frac{\partial \alpha_2}{\partial z_2}|_{z_1}
\end{array} \right)
=\left(\begin{array}{cc}
\frac{\partial \zeta}{\partial z}|_{\bar{z}} & \frac{\partial \zeta}{\partial \bar{z}}|_z \\ 
\frac{\partial \bar{\zeta}}{\partial z}|_{\bar{z}} & \frac{\partial \bar{\zeta}}{\partial \bar{z}}|_z
\end{array} \right)
= J_\eta \xC
\]
Hence using Lemma \ref{lemmalefs2} and the magnification equation (\ref{mag}), the result follows.
\end{proof}

\section{Discussion}

The fixed point formulation for gravitational lensing proposed in Definitions \ref{definlens1} and \ref{definlens2} is based on the physical theory in the given approximation since it utilizes the split of the terms due to the geometric and Shapiro time delay in the lensing map (\ref{lensmap}). Notice that the non-degeneracy condition (\ref{trans}) assumed for the $q \in \fix(f)$, which can also be expressed as $\det(I_{2m}-J_f)(q)\neq 0$, is identical to the requirement that $\mu(q)$ be finite. This is indeed the usual regularity condition for images\cite{Pe98}, which is fulfilled if and only if the source is not on a caustic.
\\
It is well known that Morse theory can be applied to gravitational lensing to yield constraints on the number of images, such as the Odd Number Theorem, and a discussion of this may be found in Petters, Levine and Wambsganss\cite{Pe98} and references therein. Now the Theorem shown here is a statement about a topological invariant of the complexified lensing map. In fact, the magnification invariant is a phenomenon inherent to the complex formulation of gravitational lensing since the real Lefschetz fixed point formula\cite{Gr78} is a sum of integer numbers which, clearly, cannot account for magnifications. Thus for the Theorem to hold in real gravitational lensing, there has to be a maximal domain where all images are real, that is, complex or \textit{spurious roots} must be absent as discussed by Hunter and Evans\cite{Hu01,Ev02}. This remains a model dependent requirement, and is the case for the point mass and Plummer models mentioned in the introduction. We now give a new astronomically motivated example where the Theorem applies.

\begin{examp}
A simple model for galaxies is the spherically symmetric modified Hubble profile\cite{Bi87} and a constant mass to light ratio. Then the mass density approximates $\rho \propto r^{-3}$ at large radii $r$. Consider a filament of such galaxies in the lens plane $L$ along the $x_2$-axis, say, such that the projected surface density is $\sigma(\mathbf{x})=\sigma_0/x_1^2$, where $\sigma_0$ is a constant of the model. Then the complexified lensing map becomes $\zeta=z+4\sigma_0/(z+\bar{z})$, so that the induced map on $\C^2$
\begin{eqnarray*}
f_1(z_1,z_2)&=&\zeta-\frac{4\sigma_0}{z_1+z_2} \\
f_2(z_1,z_2)&=&\bar{\zeta}-\frac{4\sigma_0}{z_1+z_2}
\end{eqnarray*}
is a rational lensing map in the sense of Definition \ref{definlens1}. Therefore the Theorem applies and $\sum_q \mu(q)=1$ for the complex images $q\in \fix(f)$. Now the maximal domain in the source plane $S$ is $|y_1|>\sqrt{8\sigma_0}$ where two images $q_i$ are produced, both of them real. Hence the real magnification invariant $\sum_{i=1}^2 \mu (q_i)=1$ holds.
\end{examp}

For circularly symmetric lenses, the lensing map (\ref{lensmap}) can be reduced to one variable $z$. Written as $w(z)=0$ and provided $w$ is entire, it has only real roots if and only if $w$ is in the Laguerre-P\'{o}lya class,
\[
w(z)=ce^{az-bz^2}z^n \prod_k\left(1-\frac{z}{z_k}\right)e^{z/z_k},
\]
where $a\in \R,b\in \Rp,c \in \C\backslash \{0\},n\in \Np$ and $\{z_k \}$ is a sequence of non-zero real numbers such that $\sum_k z_k^{-2}$ is finite. This was first discussed by P\'{o}lya (1913)\cite{Po74}. In general, however, the sufficient and necessary conditions for all images in the maximal domain to be real are unknown. Hence the question of what is the most general lensing map to allow real magnification invariants may be posed as a potentially interesting problem in algebraic and enumerative geometry.

\section*{ACKNOWLEDGMENTS}
I would like to thank Wyn Evans (Institute of Astronomy, Cambridge) and Gary Gibbons (Department of Applied Mathematics and Theoretical Physics, Cambridge) for invaluable discussions and gratefully acknowledge funding by the Particle Physics and Astronomy Research Council (United Kingdom).

\end{document}